\documentclass{DISproc}

\begin{document}
\title{Fragmentation Functions at Belle}

\author{{\slshape Martin Leitgab, for the Belle Collaboration}\\[1ex]
University of Illinois at Urbana-Champaign, Department of Physics, 1110 West Green Street,
Urbana, IL 61801, USA}

\contribID{64}

\doi  

\maketitle

\begin{abstract}
Fragmentation functions (FFs) describe the formation of final state particles from a partonic initial state. Precise knowledge of these functions is a key ingredient in accessing quantities such as the nucleon spin structure in semi-inclusive deep-inelastic scattering and proton proton collisions. However, fragmentation functions can currently not be determined from first principles Quantum Chromodynamics and have to be extracted from experimental data. The Belle experiment at KEK, Japan, provides a large data sample for high precision measurements on $e^{+}e^{-}$ annihilations allowing for first-time or more precise extractions of fragmentation functions. Analyses for extractions of spin-independent (unpolarized FFs) as well as spin-dependent fragmentation functions (interference FFs) at Belle are presented.
\end{abstract}


\section{Precision measurement of charged pion and kaon multiplicities}

\subsection{Motivation and outline}
A number of polarized Semi-Inclusive Deep-Inelastic Scattering (SIDIS) and polarized proton-proton scattering (pp) experiments are measuring quark and gluon distributions via QCD analysis. Unpolarized hadron fragmentation functions are input quantities for such analyses and currently limit the accuracy with which e.g. parton helicity distributions can be determined. Two recent studies have extracted hadron fragmentation functions from datasets of hadron production in $e^{+}e^{-}$~\cite{kumanopaper}, and from a combination of $e^{+}e^{-}$, SIDIS and pp datasets~\cite{marcopaper}. In both studies, the authors emphasize the need for a precision hadron multiplicity measurement from $e^{+}e^{-}$ annihilation data at low center-of-mass energies, compared to the bulk of existing datasets taken at the Large Electron-Positron Collider (LEP) at CERN. This precision dataset is expected to improve in particular the knowledge of the gluon fragmentation functions. 

Presented in this section is a precision measurement of hadron multiplicities on about $220 \times 10^{6}$ $e^{+}e^{-}$ annihilation events taken with the Belle experiment at KEK, Japan, at a center-of-mass energy of $10.52$ GeV, $60$ MeV below the $\Upsilon(4S)$ resonance. The multiplicities are measured in dependence of $z=2E_{had}/\sqrt{s}=E_{had}/E_{parton}$, the hadron energy relative to half of the center-of-mass energy, on an interval $0.2 \leq z < 1.0$. The measured hadrons are produced in reactions $e^{+}e^{-} \rightarrow q\overline{q}$, where $q~=~\{u,~d,~s,~c\}$.
 

\subsection{Experimental-data-based calibration of the Belle particle identification}

The Belle detector contains several subsystems which allow for particle identification (PID) by imposing cuts on likelihood values extracted from measurements of these subsystems. The likelihood cut selections yield fairly accurate PID but need to be calibrated in the context of a high precision measurement, such that measured hadron yields can be corrected for particle misidentification. Particle misidentification changes hadron yields up to $10 \%$ for pions and up to $20\%$ for kaons, depending on hadron momentum. The correction is performed through an unfolding technique based on inverse $5 \times 5$ PID probability matrices. 

\begin{figure}
\begin{center}
  		\begin{minipage}[t]{\linewidth}		  
				\begin{minipage}[t]{0.50\linewidth}
					\includegraphics[width=\linewidth]{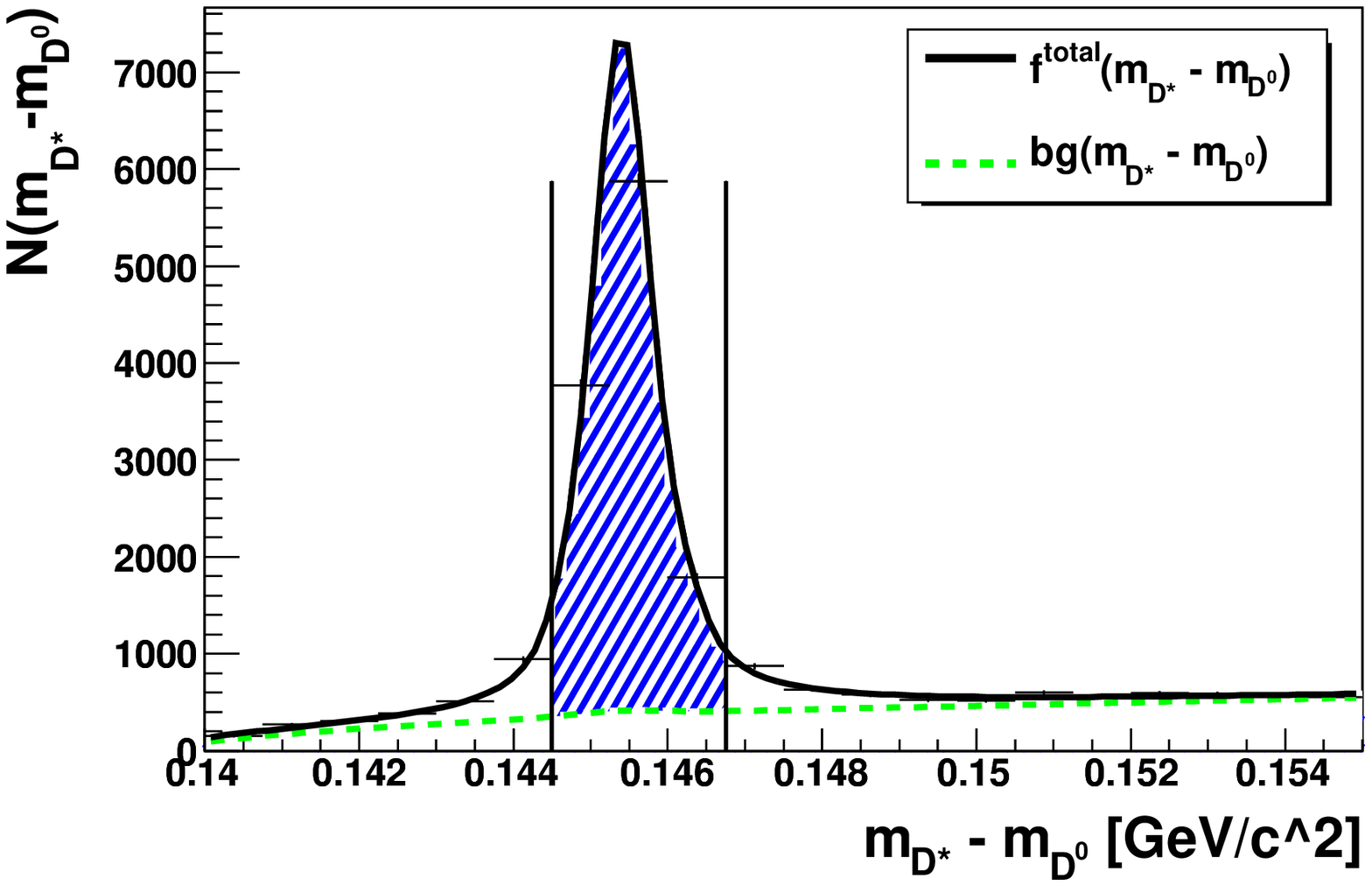}
					\begin{center} a) \end{center}
				\end{minipage}
				\hfill
				\begin{minipage}[t]{0.50\linewidth}
					\includegraphics[width=0.95\linewidth]{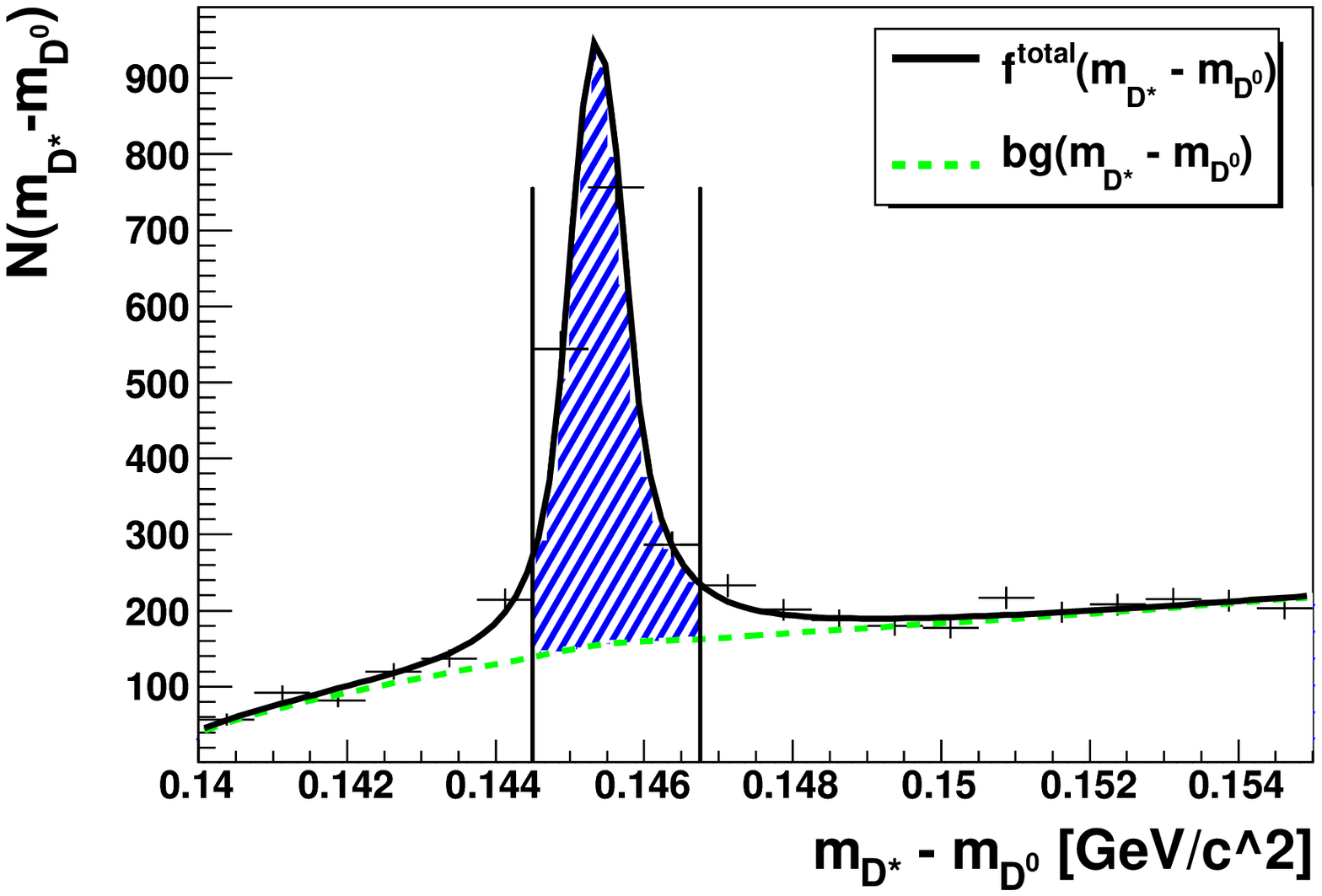}
					\begin{center} b) \end{center}
				\end{minipage}
			\end{minipage}
\end{center}
		\caption{Fits of experimental invariant mass distributions $m_{D^{*}}-m_{D^{0}}$ from decays $D^{*} \rightarrow \pi + D^{0} \rightarrow \pi + (K\pi)$. Figures a) and b) show distributions containing negatively-charged kaon candidate tracks with laboratory frame momentum $1.4 < p_{\rm lab} < 1.6$ GeV/c and laboratory frame azimuthal angle $77.9 < \theta_{\rm lab} < 89.0$ degrees. Figure b) shows all contributions to Figure a) where the kaon candidate additionally passes PID likelihood cuts to select pions. The PID misidentification of kaons as pions can be extracted from the ratio of the hatched areas, yielding the probability $p_{(K^{-} \rightarrow \pi^{-})} = 0.111 \pm 0.004$.} 
		\label{fig:pidfits}
	\end{figure}

PID probability matrices are obtained from analyzing decays of particles in which the species $i$ of the decay products can be determined through purely kinematic means. In such a sample, additional cuts on track PID likelihood variables selecting a species $j$ define a subsample of tracks. Comparing the number of cut-selected tracks to the total number of tracks in the sample allows to extract PID probabilities $p_{(i \rightarrow j)}$. Complete $5 \times 5$ matrices of PID probabilities are extracted from experimental data for species $\{i,j\} = \{ e,\mu,\pi,K,p\}$ by analyzing decays of $D^{*}$, $\Lambda$ and $J/\psi$ particles. This data-driven PID calibration avoids the dependence on the modeling of Belle PID detectors in GEANT~\cite{geant}. As an example, Figure~\ref{fig:pidfits} displays fits of invariant mass distributions $m_{D^{*}}-m_{D^{0}}$ from decays $D^{*} \rightarrow \pi + D^{0} \rightarrow \pi + (K\pi)$. The kaon candidates in contributions to Figure b) fulfill an additional PID likelihood cut to select pions. The ratio of the hatched peak areas is equal to the PID probability $p_{(K^{-} \rightarrow \pi^{-})}$.

For kinematic regions not accessible to this calibration method, an extrapolation algorithm is used to obtain PID calibration information. The PID probability matrices from all kinematic areas are inverted and then applied to the measured yields $\overrightarrow{N}_{meas} = ( N_{e}, N_{\mu}, N_{\pi}, N_{K},  N_{p})$ to obtain PID-corrected pion and kaon yields $N_{PID-corr}$. All uncertainties of the extracted PID probabilities are propagated through the inversion process using a Monte Carlo technique and are assigned to the PID-corrected yields as systematic uncertainties. 

\subsection{Other systematic corrections, preliminary results and outlook}

\begin{figure}
\centering
  		\begin{minipage}[t]{0.45\linewidth}		  
					\includegraphics[width=\linewidth]{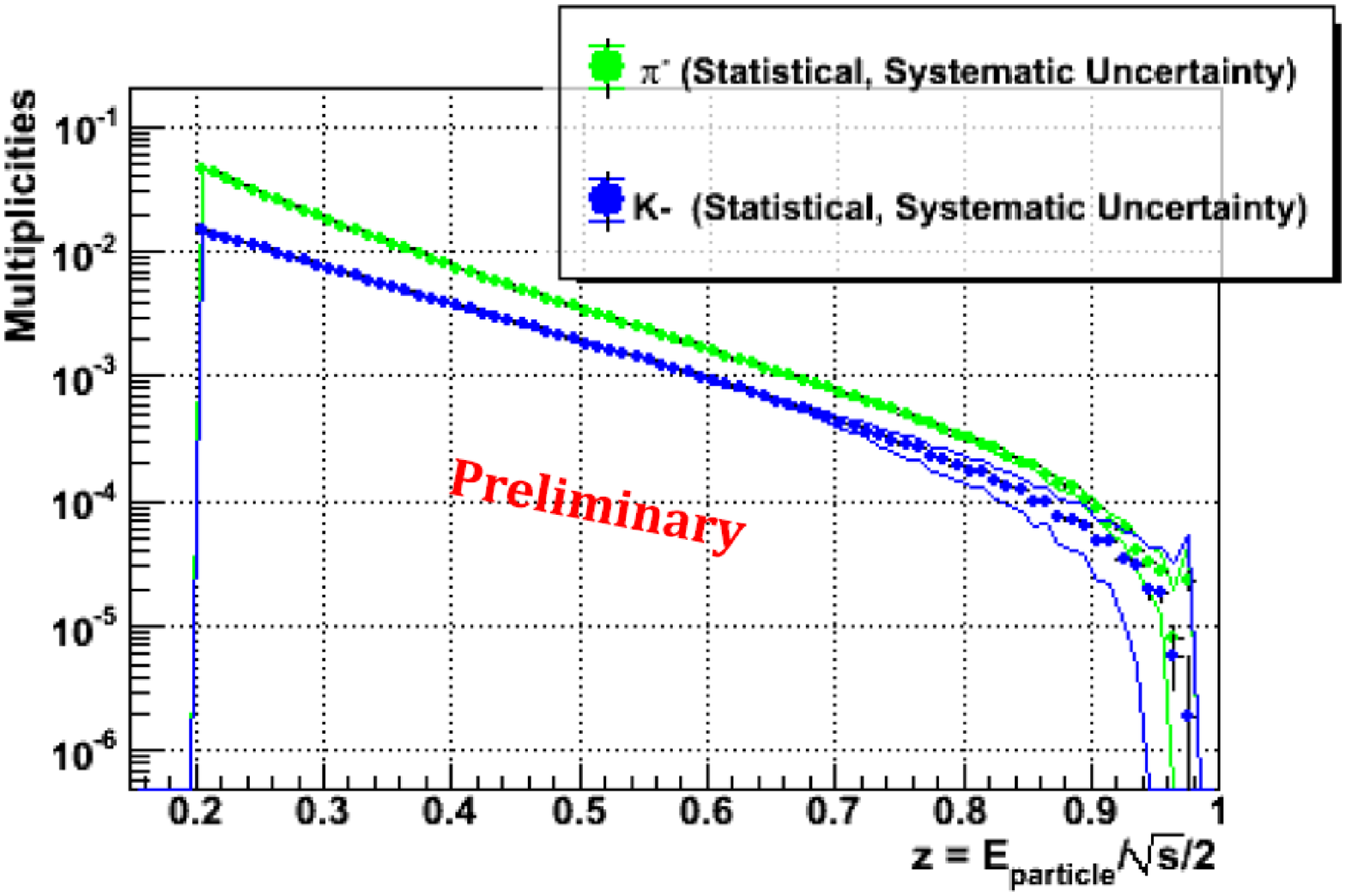}
					\begin{center} a) \end{center}
			\end{minipage}
  		\begin{minipage}[t]{0.45\linewidth}		  
					\includegraphics[width=\linewidth]{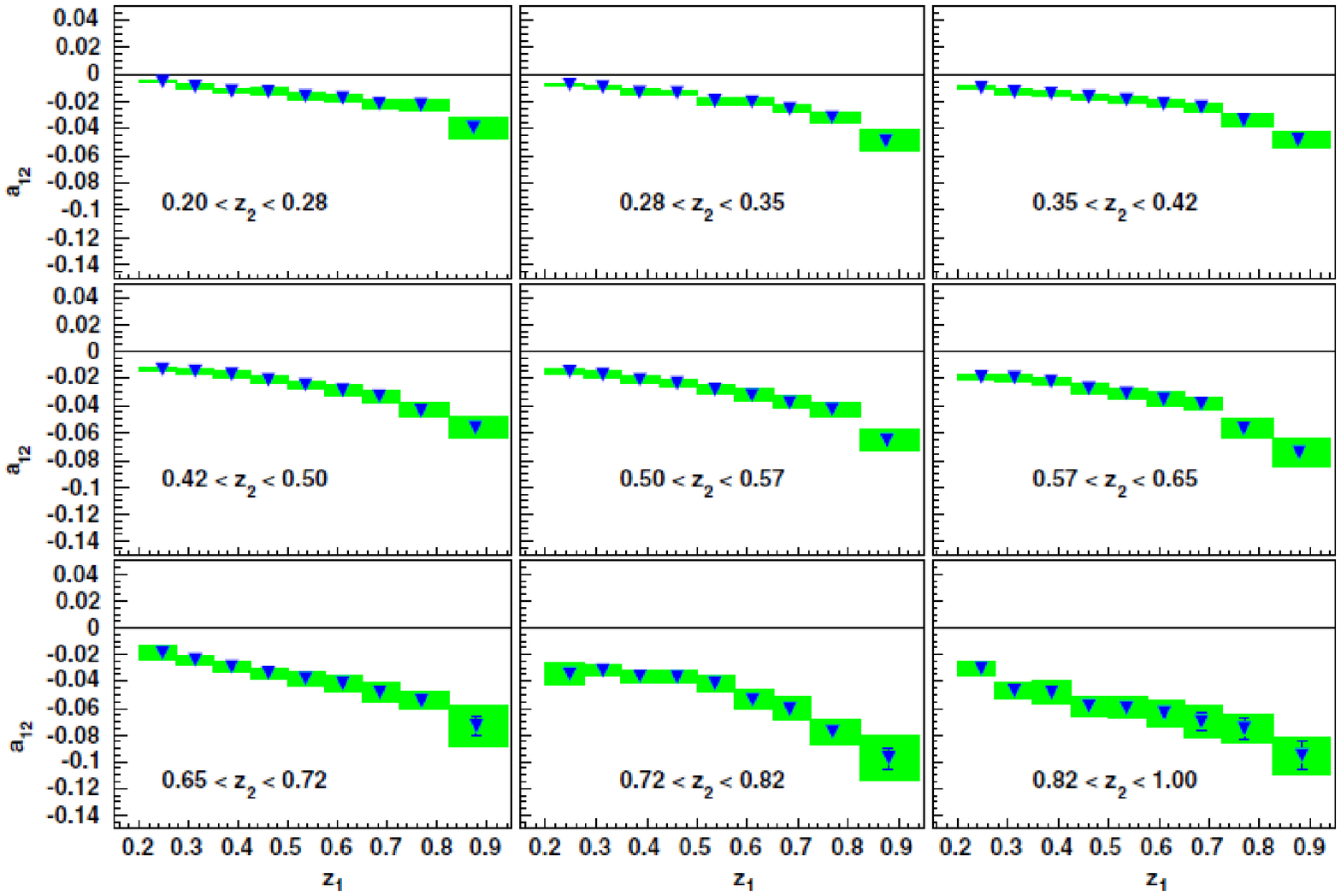}
					\begin{center} b) \end{center}
			\end{minipage}
			\caption{a) Measurement of hadron multiplicities: Preliminary negatively-charged pion and kaon multiplicities including statistical and systematic uncertainties. An additional $1.4\%$ normalization uncertainty is not shown. b) Measurement of the interference fragmentation function: Azimuthal modulations $a12$ of two-pion yields for a $9 \times 9$ $z_{1}$, $z_{2}$ binning, as a function of $z_{2}$ for all $z_{1}$ bins. Indices $1$ and $2$ refer to the respective reaction hemisphere.}
			\label{fig:collinsplanesandangledef}
	\end{figure}

\begin{figure}
\begin{center}
  		\begin{minipage}[t]{\linewidth}		  
				\begin{minipage}[t]{0.45\linewidth}
					\includegraphics[width=\linewidth]{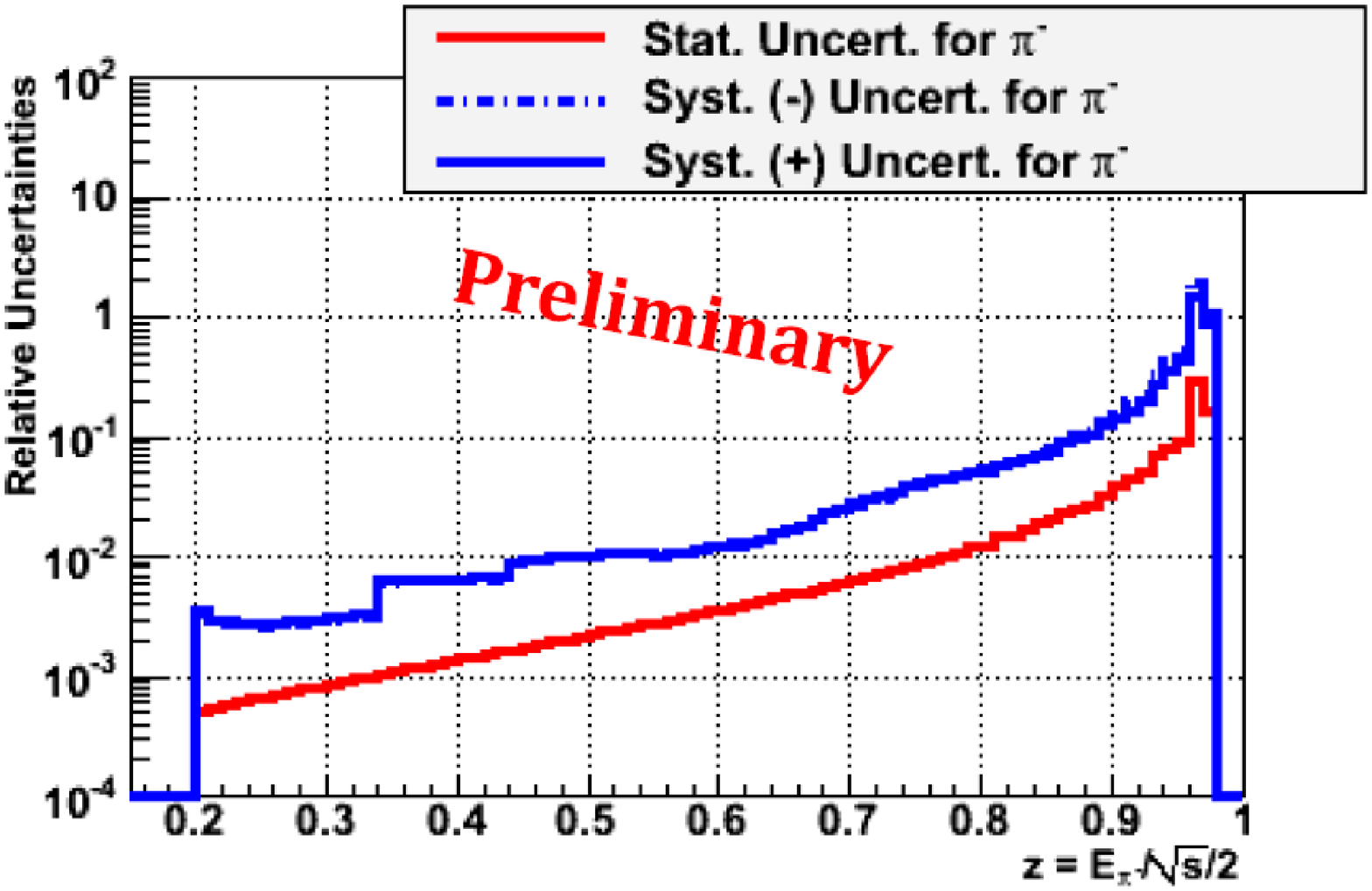}
					\begin{center} a) \end{center}
				\end{minipage}
				\hfill
				\begin{minipage}[t]{0.45\linewidth}
					\includegraphics[width=0.95\linewidth]{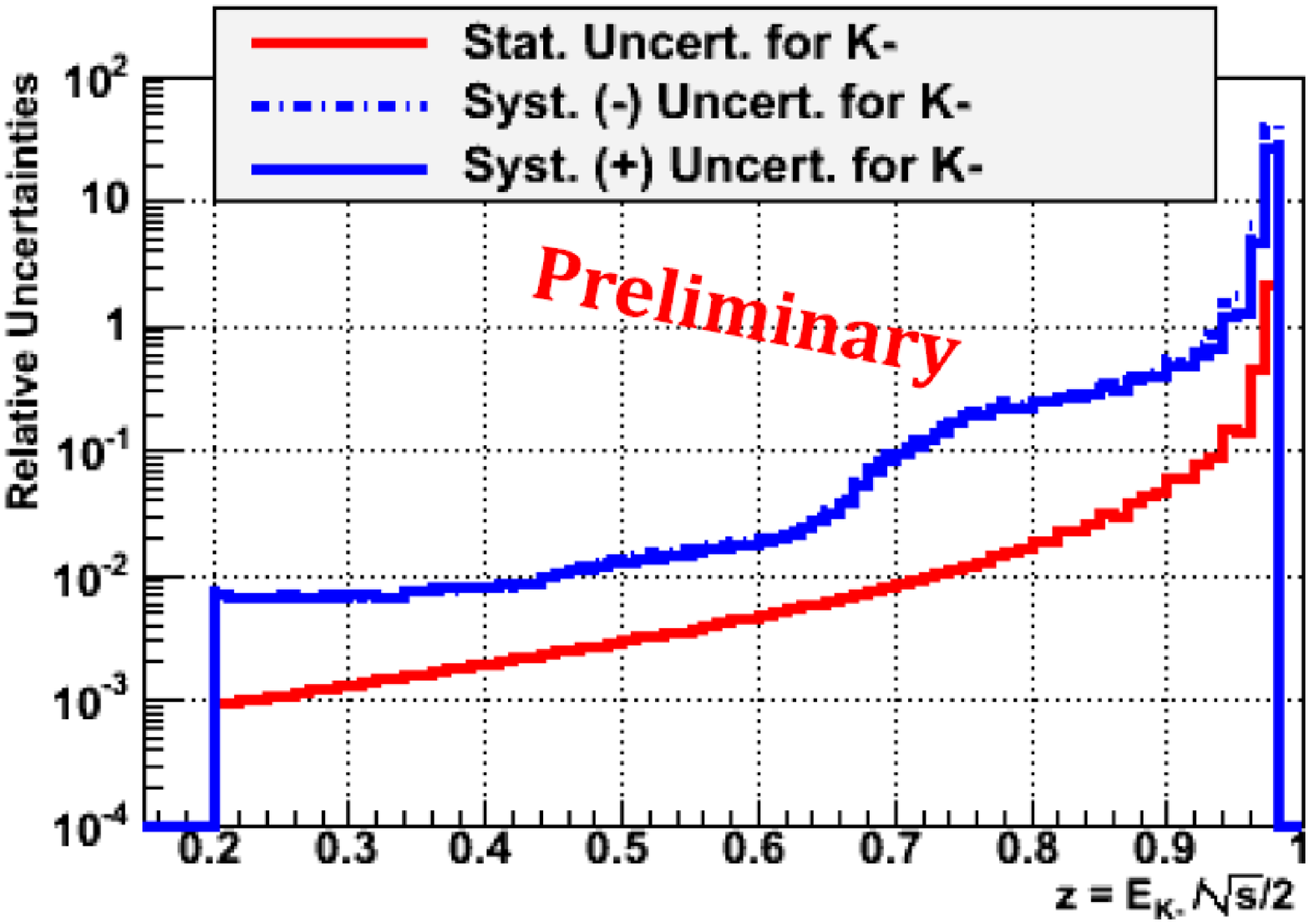}
					\begin{center} b) \end{center}
				\end{minipage}
			\end{minipage}
\end{center}
		\caption{Relative uncertainties of preliminary negatively-charged pion (a) and kaon (b) multiplicities. An additional $1.4\%$ normalization uncertainty is not shown.}
		\label{fig:prelimplotsreluncerts}
	\end{figure}

In addition to particle misidentification, the measured experimental data yields are also corrected for sample purity, kinematical smearing, decay-in-flight, detector interaction/shower particles, detector and tracking efficiencies and analysis acceptance. The presence of initial state radiation is accounted for as well.
Figure~\ref{fig:collinsplanesandangledef} a) shows preliminary negatively charged pion and kaon multiplicities including statistical and systematic uncertainties from about $220 \times 10^{6}$ $e^{+}e^{-}$ annihilations. In Figure~\ref{fig:prelimplotsreluncerts} a) and b) the relative size of preliminary statistical and systematic uncertainties is given. Final systematic uncertainties will likely remain below $2\%$ ($3\%$) for $\pi$ ($K$) spectra for $z < 0.6$, and will increase with $z$ up to $14\%$ ($50\%$) for $\pi$ ($K$) spectra through $z~\sim0.9$, respectively. Currently, the last systematic studies are being finished and timely journal submission for publication is expected.  

\section{Measurement of the interference fragmentation function for charged pion pairs at Belle}

Spin-dependent, chiral-odd fragmentation functions can be used to extract transverse spin quark distributions (so called transversity distributions) in the nucleon from polarized SIDIS and pp scattering experiments. One candidate for such a chiral-odd function is the interference fragmentation function (IFF). First proposed by Collins, Heppelman and Ladinsky~\cite{IFFCollinsref}, the IFF describes the fragmentation of a polarized quark into two hadrons correlated via partial wave interference. The corresponding observable is an azimuthal dependence in the production cross-section of hadron pairs. At Belle, the product of two IFFs can be measured by identifying one hadron pair in either hemisphere in a two-jet event $e^{+}e^{-} \rightarrow q \bar{q} \rightarrow (hh)(hh)X$. The IFF does not depend on transverse momenta and therefore its factorization and evolution can be described in a collinear approach. This makes the IFF an attractive alternative compared to the usually used Collins fragmentation function (previously measured for pions at Belle, cf. References~\cite{collinspaper1} and~\cite{collinspaper2}) for the extraction of parton transversity distributions.

Azimuthal correlations between the planes spanned by two hadron pairs and the interaction plane defined by the incoming lepton momenta and the quark-antiquark momentum axis (approximated by the thrust axis) are extracted. The raw azimuthal yields are normalized to the average hadron pair yield and fitted. This approach is not susceptible to QCD radiative effects and the $cos (\phi_{1} + \phi_{2})$ azimuthal modulation $a_{12}$ is directly proportional to the product of two IFF $H^{\angle}_{1}(z,m)$, where $z = E_{pair}/E_{parton}$ represents the fractional energy of the hadron pair and $m$ its invariant mass. Analysis results on a dataset of $672~fb^{-1}$ containing $711 \times 10^{6}$ $\pi^{+}\pi^{-}$ pairs collected at the $\Upsilon$($4$S) resonance have been published in Reference~\cite{iffref}. Figure~\ref{fig:collinsplanesandangledef} shows azimuthal modulations rising with the fractional energy of the hadron pair to significant values of up to $10\%$ for highest $z$.  

The Belle Collaboration is currently also pursuing measurements of the kaon Collins FF, the kaon and $\pi^{0}$ IFFs and the di-hadron FF. In addition, measurements are being prepared to extract the $k_{t}$ dependence of the unpolarized and the di-hadron FFs.
 

{\raggedright
\begin{footnotesize}



\end{footnotesize}
}


\end{document}